\documentclass[preprint,12pt]{elsarticle}
\usepackage{amsmath,amsfonts,enumitem,latexsym,amssymb,graphics,epsfig,subfigure,xcolor,color,makeidx,soul,multirow,placeins,float}
\usepackage[colorlinks,
            linkcolor=blue,
            anchorcolor=blue,
            citecolor=green
            ]{hyperref}
\def\ltsima{$\; \buildrel < \over \sim \;$}
\def\ltsim{\lower.5ex\hbox{\ltsima}}
\def\gtsima{$\; \buildrel > \over \sim \;$}
\def\gtsim{\lower.5ex\hbox{\gtsima}}
\def\simlt{\mathrel{\hbox{\rlap{\hbox{\lower4pt\hbox{$\sim$}}}\hbox{$<$}}}}
\def\simgt{\mathrel{\hbox{\rlap{\hbox{\lower4pt\hbox{$\sim$}}}\hbox{$>$}}}}

\def\PDE#1{partial differential equation#1 (PDE#1)\gdef\PDE{PDE}}
\def\FD#1{finite difference#1 (FD#1)\gdef\FD{FD}}
\def\pinn#1{Physics-Informed Neural Network#1 (PINN#1)\gdef\pinn{PINN}}
\def\pn#1{post-Newtonian#1 (PN#1)\gdef\pn{PN}}
\def\qnm#1{quasi-normal mode#1 (QNM#1)\gdef\qnm{QNM}}
\def\isco#1{innermost stable circular orbit#1 (ISCO#1)\gdef\isco{ISCO}}
\def\eos#1{equation#1 of state (EOS#1)\gdef\eos{EOS}}
\def\tov#1{Tolman-Oppenheimer-Volkoff#1 (TOV#1)\gdef\tov{TOV}}
\def\ns#1{neutron star#1 (NS#1)\gdef\ns{NS}}
\def\gw#1{gravitational wave#1 (GW#1)\gdef\gw{GW}}
\def\WD#1{white dwarf#1 (WD#1)\gdef\WD{WD}}
\def\wdbh#1{white dwarf -- black hole#1 (WDBH#1)\gdef\wdbh{WDBH}}
\def\bbh#1{binary black holes#1 (BBH#1)\gdef\bbh{BBH}}
\def\bhns#1{black hole -- neutron star#1 (BHNS#1)\gdef\bhns{BHNS}}
\def\nsns#1{neutron star -- neutron star#1 (NSNS#1)\gdef\nsns{NSNS}}
\def\emri#1{Extreme Mass-Ratio Inspiral#1 (EMRI#1)\gdef\emri{EMRI}}
\def\emrb#1{Extreme Mass-Ratio Binaries#1 (EMRB#1)\gdef\emrb{EMRB}}
\def\sgrb#1{short gamma-ray burst#1 (SGRB#1)\gdef\sgrb{SGRB}}
\def\grb#1{gamma-ray burst#1 (GRB#1)\gdef\grb{GRB}}
\def\imbh#1{intermediate mass black hole#1 (IMBH#1)\gdef\imbh{IMBH}}
\def\smbh#1{supermassive black hole#1 (SMBH#1)\gdef\smbh{SMBH}}
\def\bh#1{black hole#1 (BH#1)\gdef\bh{BH}}
\def\ulx#1{ultra-luminous x-ray source#1 (ULX#1)\gdef\ulx{ULX}}
\def\lmxbs{low-mass x-ray Binaries (LMXBs)\gdef\lmxbs{LMXBs}\gdef\lmxb{LMXB}}
\def\lmxb{low-mass x-ray Binary (LMXB)\gdef\lmxbs{LMXBs}\gdef\lmxb{LMXB}}
\def\dns#1{double neutron star#1 (DNS#1)\gdef\dns{DNS}}
\def\mots#1{marginally outer trapped surface#1 (MOTS#1)\gdef\mots{MOTS}}

\newcommand\mathnew{\mathsurround=0pt}

\def\simov#1#2{\lower .5pt\vbox{\baselineskip0pt \lineskip-.5pt
        \ialign{$\mathnew#1\hfil##\hfil$\crcr#2\crcr\sim\crcr}}}
\def\MPR#1{{\it Moving Puncture Recipe}#1 (MPR#1)\gdef\MPR{MPR}}
\def\bh#1{black hole#1 (BH#1)\gdef\bh{BH}}
\def\ahz#1{apparent horizon#1 (AH#1)\gdef\ahz{AH}}
\def\bbh#1{binary black hole#1 (BBH#1)\gdef\bbh{BBH}}
\def\qnm#1{quasi-normal mode#1 (QNM#1)\gdef\qnm{QNM}}
\def\isco#1{innermost stable circular orbit#1 (ISCO#1)\gdef\isco{ISCO}}

\def\tde#1{tidal disruption event#1 (TDE#1)\gdef\tde{TDE}}
\def\whd#1{white dwarf#1 (WD#1)\gdef\whd{WD}}

\journal{Journal of Computational Physics}

\begin{document}

\title{Calculating Quasi-Normal Modes of Schwarzschild Black Holes with Physics Informed Neural Networks}

\author{Nirmal Patel,
		Aycin Aykutalp,
        Pablo Laguna
}

\affiliation{organization={Center of Gravitational Physics and Department of Physics},
           addressline={The University of Texas at Austin}, 
           city={Austin},
           state={TX},
           postcode={78712}, 
           country={USA}}

\begin{abstract}
Machine learning, particularly neural networks, has rapidly permeated most activities and work where data has a story to tell. Recently, deep learning has started to be used for solving differential equations with input from physics, also known as \pinn{s}. We present a study showing the efficacy of \pinn{s} for solving the Zerilli and the Regge-Wheeler equations in the time domain to calculate the quasi-normal oscillation modes of a Schwarzschild black hole. We compare the extracted modes with those obtained with finite difference methods. 
Although the \pinn{} results are competitive, with a few percent differences in the quasi-normal modes estimates relative to those computed with finite difference methods, the real power of \pinn{s} will emerge when applied to large dimensionality problems. 
\end{abstract}



\maketitle

\section{Introduction}\label{scheme1}
Deep learning has grown exponentially in the last couple of decades, with an astonishing range of applications in computer vision, natural language processing, autonomous vehicles, and many more. Deep learning has also made its way into the scientific fields that deal with complex dynamic systems involving various input variables using automatic differentiation \cite{baydin}. As Raissi et al. \cite{RAISSI2019686} have pointed out, feeding physics to the neural network improves its accuracy while making it data-efficient. They coined the term ``Physics Informed Neural Networks". There is strong evidence that \pinn{s} could potentially provide a new insight to tackle the \emph{curse of dimensionality}, as introduced by Bellman \cite{Bellman}, which is still a hindrance for current numerical methods. 

\pinn{s} have been recently utilized in gravitational physics problems. They have been used to solve Riemann problems applied to relativistic hydrodynamics \cite{font}, to compute the \qnm{s} of black holes in de Rham-Gabadadze-Tolley massive gravity, but in this case, the computation was carried out in the frequency domain \cite{haddou2023quasinormal}. 

In this work, we present a study showing the efficacy of \pinn{s} for solving the Zerilli and the Regge-Wheeler equations in the time domain to compute the \qnm{s} of a Schwarzschild \bh{}. We compare the extracted modes with those obtained with \FD{} methods. We also show the accuracy of \pinn{s} for this problem in terms of its scaling with the number of nodes and layers. We found that the \pinn{s} results are competitive, with a few percent differences in the \qnm{s} relative to those with \FD{} methods. 

The paper is organized as follows. In Sec.~\ref{sec:nn}, we present a brief description of neural networks. The description is extended to \pinn{s} in Sec.~\ref{sec:pinns}. In Sec.~\ref{sec:Perturbations}, we briefly summarize the equations governing the perturbation of a Schwarzschild \bh{} and its \qnm{s}. In Sec.~\ref{sec:Arch}, we describe the \FD{} methodology and \pinn{} architecture used to solve the Zerilli and Regge-Wheeler equations. Results are given in Sec.~\ref{Results}, with conclusions in Sec.~\ref{Conclusions}.

\section{Neural Networks: Fundamentals and Application in PINNs}\label{sec:nn}
Neural networks are computational frameworks inspired by the biological neural networks of our brains \cite{mcculloch1943logical}. At their core, they consist of layers of nodes or neurons interconnected in a way that allows for complex data processing and pattern recognition. This makes them particularly suited for tasks ranging from simple classification to complex physics problems.

Neural networks consist of an input layer, multiple hidden layers, and an output layer. Each layer contains neurons that process incoming data. Early neural networks used simple units called Threshold Logic Units (TLUs, \cite{rosenblatt1958perceptron}). Modern networks use more advanced artificial neurons, which apply continuous, differentiable activation functions, allowing for more complex and nuanced data modeling.

Learning in a neural network is a systematic process involving several interconnected steps. These steps enable the network to adjust its internal parameters (weights and biases) based on the data it processes, thereby improving its performance in tasks such as classification or regression.

\begin{itemize}\label{NNs}
\item \textbf{Initialization}: Before learning begins, the neural network's weights and biases are initialized. This can be done randomly or using specific initialization techniques like Xavier or He initialization. Initialization is crucial as it affects the convergence speed and the ability of the network to reach an optimal solution \citep{pmlr-v9-glorot10a}.
\item \textbf{Forward Propagation} Data flows from the input through the hidden layers to the output, with each neuron computing a weighted sum of its inputs followed by an activation function \cite{Goodfellow-et-al-2016}.
\begin{itemize}
\item{\emph{Input Layer}: The learning process begins with input data being fed into the network. Each neuron in the input layer represents a feature of the input data.}
\item{\emph{Hidden Layers}: The data then passes through one or more hidden layers. In each layer, the neurons perform a weighted sum of their inputs, followed by an activation function. The activation function introduces non-linearity, allowing the network to learn complex patterns.}
\item{\emph{Output Layer}: The final layer produces the network's output. The nature of this output depends on the task (e.g., a single value for regression, probability scores for classification)}.
\end{itemize}

\item{\textbf{Loss Calculation:} A loss function quantifies the error between the network’s predictions and actual data, crucial for guiding the learning process. Once the network produces an output, the loss function calculates the error. This error is the difference between the network's prediction and the actual target values. Common loss functions include Mean Squared Error (MSE) for regression tasks and Cross-Entropy for classification tasks.
}
\item{\textbf{Backpropagation}: Backpropagation is the process through which the network learns \cite{rumelhart1986learning}. The goal is to minimize the loss by adjusting the weights and biases. The network computes the gradient of the loss function, propagating this information back through the network to inform how weights should be adjusted. It involves computing the gradient of the loss function with respect to each weight and bias in the network. This computation uses the chain rule of calculus, a method known as the gradient descent. The gradients indicate how much and in which direction the weights and biases should be adjusted to reduce the loss.}
\item{\textbf{Weight Update}: Using the gradients calculated during backpropagation, the weights and biases are updated \cite{kingma2017adam}. This step is crucial for the learning process. The magnitude of the update is determined by the learning rate, a hyperparameter that dictates how big a step should be taken towards the minimum of the loss function. Optimizers like SGD (Stochastic Gradient Descent), Adam, or RMSprop are used to perform this update. They differ in how they use the gradient information to adjust the weights (e.g., some use momentum, others adjust learning rates adaptively).}
\end{itemize}

The output from forward propagation is used to calculate how far off the predictions are from the actual values. The loss guides how the model should adjust its parameters. It's the starting point for backpropagation, determining the direction for the gradient descent. The gradients from backpropagation are used directly to update the weights in a direction that minimizes the loss. Updated weights affect the next round of forward propagation, leading to different outputs and a reduced loss. The entire process (forward propagation, loss calculation, backpropagation, weight update) is repeated for a number of iterations or epochs over the training dataset. With each iteration, the network's predictions become more accurate, and the loss decrease. The learning continues until the loss converges to a minimum value, a pre-defined number of epochs is reached, or another stopping criterion is met. This cyclic process enables the neural network to learn from data, gradually improving its predictions through successive iterations.

In the context of \pinn{s}, neural networks are tailored to solve physics problems, particularly involving differential equations. \pinn{s} integrate physical laws into the learning process, improving the network's ability to generalize and learn from limited data. This integration is achieved by constructing a loss function that not only minimizes the difference between predicted and actual outputs but also ensures compliance with the physical laws governing the system.
For instance, when applied to an initial-boundary value problem, \pinn{s} aim to construct an approximate solution to a given partial differential equation (\PDE{}), respecting initial and boundary conditions. This is accomplished by minimizing a loss function that includes the residuals of the \PDE{} and these conditions. The network uses layers (denoted as L dense layers) to approximate the solution, with each layer involving weight matrices, bias vectors, and activation functions.

\section{Physics Informed Neural Networks}\label{sec:pinns}

When applied to an initial-boundary value problem, the \pinn{} method aims at constructing an approximate solution  to a \PDE{}, given initial and boundary conditions, by minimizing a loss function that involves the residuals of the \PDE{} and the initial and boundary conditions. 

Specifically, we consider the following hyperbolic \PDE{} for the function $\Phi(x,t)$
\begin{equation}\label{HyperbolicPDE}
    A\frac{\partial^2 \Phi}{\partial t^2} + 2B\frac{\partial^2 \Phi}{\partial t \partial x} + C\frac{\partial^2 \Phi}{\partial x^2} + \mathbb{D}[\Phi] = 0,
\end{equation}
in a domain $X = [x_{min}, x_{max}] \times [t_{min}, t_{max}]$. Here, $\mathbb{D}$ is a first-order differential operator, and the coefficients, in general, depend on $x$ and satisfy $B^2 - AC > 0$. The values of $\Phi(x,t_{min})$ and $\frac{\partial \Phi}{\partial t}(x,t_{min})$ are provided as initial conditions. For boundary conditions, we impose $\mathbb{L}[\Phi]=0$ at left boundary $(x_{min},t)$ and $\mathbb{R}[\Phi]=0$ at right boundary $(x_{max},t)$, where $\mathbb{L}$ and $\mathbb{R}$ are first order differential operators. 

The \pinn{} approximated solution using $L$ dense layers will be 
\begin{equation}\label{PINNApprox}
    \Phi_{\boldsymbol{\theta}}(X) = \mathbf{W}^{L}\sigma^{L}(...\mathbf{W}^{1}\sigma^{1}(\mathbf{W}^{0}X + \mathbf{b}^{0})+\mathbf{b}^{1}...) + \mathbf{b}^{L},
\end{equation}
where $\boldsymbol{\theta}$ is the trainable parameter vector, $\mathbf{W}^{i}$ are weight matrices, $\mathbf{b}^{i}$ are bias vectors, and $\sigma^i$ are activation functions.

The \PDE{} residual of the \pinn{} approximate is 
\begin{equation}\label{PINNApproxRes}
    r_{\boldsymbol{\theta}} = A\frac{\partial^2 \Phi_{\boldsymbol{\theta}}}{\partial t^2} + 2B\frac{\partial^2 \Phi_{\boldsymbol{\theta}}}{\partial t \partial x} + C\frac{\partial^2 \Phi_{\boldsymbol{\theta}}}{\partial x^2} + \mathbb{D}[\Phi_{\boldsymbol{\theta}}].
\end{equation}
Thus, the corresponding loss function is  
\begin{equation}\label{RLoss}
    \mathcal{L}_r(\boldsymbol{\theta}) = \sum_{n_r}
    \frac{\begin{Vmatrix}
        r_{\boldsymbol{\theta}}(x,t),
    \end{Vmatrix}_2^2}{|N_r|},
\end{equation}
where the sum is over $N_r$ discrete points in the interior of the domain $X$, and $\begin{Vmatrix}\,.\,\end{Vmatrix}_2$ donates L2 norm. 

The accuracy of \pinn{} can be increased by adding derivatives of $r_{\boldsymbol{\theta}}$. Such \pinn{} is known as ``Gradient-enhanced \pinn{}" \cite{YU2022114823}. The loss functions involving the derivatives of residuals are 
\begin{equation}\label{RXLoss}
    \mathcal{L}_{r_x}(\boldsymbol{\theta}) = \sum_{n_r}
    \frac{\begin{Vmatrix}
        \frac{\partial r_{\boldsymbol{\theta}}}{\partial x}(x,t)
    \end{Vmatrix}_2^2}{|N_r|},
\end{equation}
and
\begin{equation}\label{RTLoss}
    \mathcal{L}_{r_t}(\boldsymbol{\theta}) = \sum_{n_r}
    \frac{\begin{Vmatrix}
        \frac{\partial r_{\boldsymbol{\theta}}}{\partial t}(x,t)
    \end{Vmatrix}_2^2}{|N_r|}. 
\end{equation}
The loss function for the initial conditions are 
\begin{equation}\label{ICLoss}
    \mathcal{L}_{ic}(\boldsymbol{\theta}) = \sum_{ n_{i}}
    \frac{\begin{Vmatrix}
        \Phi_{\boldsymbol{\theta}}(x,t_{min}) - \Phi(x,t_{min})
    \end{Vmatrix}_2^2}{|N_{i}|},
\end{equation}
and 
\begin{equation}\label{IVLoss}
    \mathcal{L}_{iv}(\boldsymbol{\theta}) = \sum_{ n_{i}}
    \frac{\begin{Vmatrix}
        \frac{\partial \Phi_{\boldsymbol{\theta}}}{\partial t}(x,t_{min}) - \frac{\partial \Phi}{\partial t}(x,t_{min})
    \end{Vmatrix}_2^2}{|N_{i}|}\,,
\end{equation}
where the sums are over $N_i$ discrete points $(x,t_{min}) \in X$.
And, at the left and right boundaries, the loss terms read
\begin{equation}\label{BLLoss}
    \mathcal{L}_{bl}(\boldsymbol{\theta}) = \sum_{ n_{b}}
    \frac{\begin{Vmatrix}
        \mathbb{L}[\Phi_{\boldsymbol{\theta}}]
    \end{Vmatrix}_2^2}{|N_{b}|},
\end{equation}
and
\begin{equation}\label{BRLoss}
    \mathcal{L}_{br}(\boldsymbol{\theta}) = \sum_{ n_{b}}
    \frac{\begin{Vmatrix}
        \mathbb{R}[\Phi_{\boldsymbol{\theta}}]
    \end{Vmatrix}_2^2}{|N_{b}|},
\end{equation}
where the sums are over $N_{b}$ points $(x_{min},t) \in X$ for the sum in Eq.~(\ref{BLLoss}) and 
$N_{b}$ points $(x_{max},t) \in X$ for the sum in Eq.~(\ref{BRLoss}).

Finally, the vector of loss functions will be
\begin{equation}\label{LossVector}
    \begin{split}
        \boldsymbol{\mathcal{L}}(\boldsymbol{\theta})= & [\mathcal{L}_{r}(\boldsymbol{\theta}), \mathcal{L}_{r_x}(\boldsymbol{\theta}), \mathcal{L}_{r_t}(\boldsymbol{\theta}), \mathcal{L}_{ic}(\boldsymbol{\theta}), \\
        & \mathcal{L}_{iv}(\boldsymbol{\theta}), \mathcal{L}_{bl}(\boldsymbol{\theta}), \mathcal{L}_{br}(\boldsymbol{\theta})]\,.
    \end{split}
\end{equation} 
Thus, given the weight vector 
\begin{equation}\label{WeightVector}
    \boldsymbol{\lambda}=[\lambda_{r}, \lambda_{r_x}, \lambda_{r_t}, \lambda_{ic}, \lambda_{iv}, \lambda_{bl}, \lambda_{br}]\,,
\end{equation} 
the loss function to be minimized would be 
\begin{equation}\label{TotLoss}
    \mathcal{L}(\boldsymbol{\theta}) = \boldsymbol{\lambda} \cdot \boldsymbol{\mathcal{L}}(\boldsymbol{\theta}). 
\end{equation}
The weight parameters are chosen so that the components of the loss function vector (\ref{LossVector}) have comparable values to minimize bias.

 The motivation for using \pinn{s} to solve \PDE{s} is because of their potential advantage for problems with large computational complexity. For a \PDE{} with $d_{i}$ dimensional input and single dimensional output, the complexity of a \pinn{} with $L$ homogeneous dense hidden layers with $n$ nodes, $N_{d}$ data points, and $N_{it}$ iterations will be $\mathcal{O}(N_{d}N_{it}(n(d_{in} + 1) + n^2(L-1)))$ \cite{freire2022computational}. 

\section{Gravitational Perturbations of Schwarzschild Black Holes}\label{sec:Perturbations}

When a \bh{} is slightly perturbed, the gravitational radiation it emits as it settles down (ring-down) has characteristic frequencies and decay times independent of the process that gave rise to the \gw{s}. The emission is in the form of \qnm{s}. The modes are a family of exponentially damped sinusoidals with frequencies and decaying times directly related to the ``hairs" of the \bh{}, namely its mass, spin, and charge~\cite{Chandrasekhar_1975}. 

We will focus only on a non-spinning and uncharged \bh{,} also called Schwarzschild \bh{}, in which the only hair in the \bh{} is its mass $M$. To study the perturbations of a Schwarzschild \bh{}, one starts with the background space-time metric 
\begin{eqnarray}
    ds^2 &=& g^0_{\mu\nu}dx^\mu dx^\nu 
    =-\left(1 - \frac{2\,M}{r}\right)dt^2\\ \nonumber
    &+& \left(1 - \frac{2\,M}{r}\right)^{-1}dt^2 + r^2\,d\Omega^2\,,
\end{eqnarray}
where  $d\Omega^2 = d\theta^2+\sin^2{\theta}\,d\varphi^2$.  One then adds a perturbation of the form $g_{\mu\nu} = g^0_{\mu\nu} + h_{\mu\nu}$, and the Einstein equations reduce to a set of linear partial differential equations for $h_{\mu\nu}$. After decomposing $h_{\mu\nu}$ in tensor spherical harmonics and with a clever combination of components of $h_{\mu\nu}$, one arrives at a single master equation for a scalar quantity $\Phi$ decomposed as~\cite{Kokkotas1999-rw}
\begin{equation}\label{eq:master}
    \Phi(t,r,\theta,\varphi) = \frac{1}{r} \sum_{\ell m} \Phi_{\ell m}(t,r)\,Y_{\ell m}(\theta,\varphi)\,.
\end{equation}
Since we only consider a spherically symmetric background, the perturbation will not depend on $\varphi$, and $m$ can be ignored. The master equation for each $\Phi_\ell$ reads
\begin{equation}\label{eq:PDE}
    -\frac{\partial^2}{\partial t^2}\Phi_\ell + \frac{\partial^2}{\partial x^2}\Phi_\ell + V_{\ell}(r)\Phi_\ell = 0,
\end{equation}
where $x$ is the ``tortoise" radial coordinate defined within  $r\in(2M, \infty)$ by 
\begin{equation}\label{Tortoise}
    x = r + 2\,M\ln\left(\frac{r}{2M} -1\right)\,;
\end{equation}
thus, $x \in(-\infty, \infty)$. For simplicity, we will drop the $\ell$ label in $\Phi$.

If the interest is with ``axial" or odd-parity perturbations, namely those that transform under a parity operation as $(-1)^{\ell+1}$, the potential $V(r)$ is given by
\begin{equation}\label{Regge-Wheeler}
            V(r) = \left(1 - \frac{2M}{r}\right)\left[\frac{2(n+1)}{r^2} + \frac{2 (1-s^2) M}{r^3}\right]\,,
\end{equation}
with $2\,n = (\ell-1)(\ell+2)$ and $s = 0, 1$, and 2 for scalar, electromagnetic, and gravitational perturbations, respectively. For our case, we set $s = 2$. The master equation, in this case, is known as the Regge-Wheeler equation. 

If, on the other hand, one is interested in ``polar" or even-parity perturbations that transform under a parity operation as $(-1)^\ell$, the potential $V$ is given by
    \begin{eqnarray}\label{Zerilli}
            V(r) &=& \left(1 - \frac{2M}{r}\right)\frac{1}{r^3(nr+3M)^2}\\ \nonumber
            &&[2\,n^2(n+1)r^3 + 6\,n^2M\,r^2 + 18\,n\,M^2r + 18M^3]
    \end{eqnarray}
For this case, the master equation is called the Zerilli equation.

The potentials have a barrier around $x \sim 1$. They also vanish as one approaches the \bh{} horizon, $r \rightarrow 2\, M$ or $x \rightarrow -\infty$, and far away from the hole, i.e., when $r$ and $x \rightarrow \infty$. A transformation connects the Zerilli and Regge-Wheeler equations~\cite{Chandrasekhar_1975}. Therefore, both equations yield the same \qnm{s}. The Zerilli equation was originally solved in the frequency domain to compute \qnm{s}~\cite{Chandrasekhar_1975}. Table~\ref{table:QNM} shows the frequencies $\omega$ and decay times $\tau$ for the most dominant $\ell$ modes in which $\Phi \propto e^{-t/\tau}\,\sin{(\omega\,t)}$.

\begin{table}
\centering
    \begin{tabular}{|c|c|c|}
        \hline
        $\ell$ & $\omega\,M$ & $\tau/M$ \\ \hline
       2& 0.37367  & 11.241 \\ 
        3& 0.59944  &10.787\\ 
     4 & 0.80918 & 10.620 \\ \hline
    \end{tabular}
    \caption{\label{table:QNM}\qnm{} parameters of a Schwarzschild \bh{} in which $\Phi_\ell \propto e^{-t/\tau}\,\sin{(\omega\,t)}$}
\end{table}

We will solve both the Zerilli and Regge-Wheeler in the time domain with both \pinn{s} and for comparison with \FD{} methods. As in the frequency domain case, we require that the perturbations are outgoing into the \bh{} horizon and outgoing at infinity~\cite{Chandrasekhar_1975}. Since the potential vanishes at the horizon and spatial infinity, the  master equation Eq.~(\ref{eq:master}) reduces to a wave equation, and we impose the following  boundary conditions:
\begin{equation}\label{BL}
    \left(\frac{\partial}{\partial t} - \frac{\partial}{\partial x}\right)\Phi = 0,
\end{equation}
as $x\rightarrow -\infty$, i.e. into the \bh{,} and
\begin{equation}\label{BR}
    \left(\frac{\partial}{\partial t} + \frac{\partial}{\partial x}\right)\Phi = 0.
\end{equation}
as $x \rightarrow \infty$. We choose as initial conditions
\begin{eqnarray}\label{IC}
    \Phi(x) &=& A \exp{\left[-\frac{(x - x_0)^2}{\sigma^2}\right]}\\
\label{IV}
    \frac{\partial \Phi}{\partial t}(x) &=& 2\frac{(x - x_0)^2}{\sigma^2}\Phi(x)\,.
\end{eqnarray}
That is, we initially have a Gaussian outgoing pulse with amplitude $A$,  width $\sigma$, and centered at $x_0$.  We set $A = 1$, $x_0 = 4\,M$, and $\sigma = 5\,M$. All quantities are reported in units of $M$.

\FloatBarrier
\section{Computational Methodology}\label{sec:Arch}

For solutions obtained with \FD{} methods, we use a uniform mesh with $N_x = 1,000$ grid points in a domain  $x/M \in [-50, 150]$, i.e., grid-spacing $\Delta x = 0.2\, M$. Spatial differentiation is approximated with 2nd-order \FD{} operators. We use a method of lines with time updates via a 4th-order Runge-Kutta method with a Courant factor of $0.5$. The time-step is then $\Delta t = 0.5\,\Delta X = 0.1\,M$. We evolve for a time span $t/M \in [0, 50]$. Thus, we take $N_t = 500$ steps, which translates into $N_{F} = N_x N_t = 5\times 10^5$ space-time grid points.

For the \pinn{s} solutions, we use $N_i=800$, $N_b=400$, and $N_r=32,000$, which translates into $N_{P} = 2 N_b+N_i+N_r = 33,600$ space-time points where the residuals are evaluated. The input nodes are $x$ and $t$, with $\Phi_{\boldsymbol{\theta}}$ the output node. To prevent PINN from adapting a blowing-up solution, we enforce a bound by applying the output transformation $\Phi_{\boldsymbol{\theta}} = A \tanh(\Phi_{\boldsymbol{\theta}})$ as a constraint. The network has four hidden layers with widths [80, 40, 20, 10]. The number of trainable parameters are $dim(\boldsymbol{\theta}) = 2 \cdot 80 + 80 + 80 \cdot 40 + 40 + 40 \cdot 20 + 20 + 20 \cdot 10 + 10 + 10 \cdot 1 + 1 = 4,521$.
 We use a Glorot uniform initializer and $\tanh{}$ activation throughout the network. The \pinn{} is trained using the ADAM optimizer (10,000 iterations) followed by the L-BFGS optimizer (15,000 iterations). For the weight vector  $\boldsymbol{\lambda}$ given in Eq. \ref{WeightVector}, we use [100, 100, 100, 1, 100, 1, 1]. The weight $\lambda_{iv} = 100$ is  because of the phase problem that will be discussed in the next section. Lastly, we use DeepXDE's residual points re-sampling feature to get a fresh batch of training points after every 100 iterations \cite{lu2021deepxde}. 

\section{Results}\label{Results}

We used three metrics for analyzing errors and deviations between \pinn{s} and \FD{} results: Root-Mean-Squared-Deviation (RMSd):
\begin{equation}\label{eqn:rms}
    \textnormal{RMSD} = \sqrt{\frac{\sum_{N_F}(\Phi - \Phi_{\boldsymbol{\theta}})^2}{N_F}},
\end{equation}
Mean-Absolute-Deviation (MAD):
\begin{equation}\label{eqn:mad}
    \textnormal{MAD} = \frac{1}{N_F}\sum_{N_F}|\Phi - \Phi_{\boldsymbol{\theta}}|,
\end{equation}
and Relative $L^2$ (RL2) norm:
\begin{equation}\label{eqn:rl2}
    \textnormal{RL2} = \sqrt{\frac{\sum_{N_F}(\Phi - \Phi_{\boldsymbol{\theta}})^2}{\sum_{N_F}\Phi^2}},
\end{equation}
where the functions are evaluated at the $N_F$ space-time grid points used to construct the \FD{} solution. The values of these metrics are shown in Table~\ref{ResultTable}. It is evident from the values of RMSD and MAD that the PINN method produces results close to those from \FD{} methods. The Relative $L^2$ error is high because most data points are zero. The formula of Relative $L^2$ error is 

\begin{table}[htbp]
    \centering
    \begin{tabular}{|c|c|c|}
        \hline
                      & \textbf{Zerilli} & \textbf{Regge-Wheeler} \\ \hline
        \textbf{RMSD} & $\ 0.0370\ $  & $\ 0.0333\ $        \\ \hline
        \textbf{MAD } & $\ 0.0182\ $  & $\ 0.0187\ $        \\ \hline
        \textbf{RL2}  & $\ 0.2806\ $  & $\ 0.2359\ $        \\ \hline
    \end{tabular}
    \caption{\label{ResultTable}RMSD, MAD, and RL2 deviations between the \FD{} and \pinn{} solutions.}
\end{table}

It is evident from Table~\ref{ResultTable} that \pinn{} results are very close to those from \FD{}, with low RMSD and MAD. The reason why the RL2 values seem high is because a significant fraction of the data points in the computational domain are zero and do not contribute to the denominator.

\begin{figure}[hbt!]
    \includegraphics[width=1\textwidth]{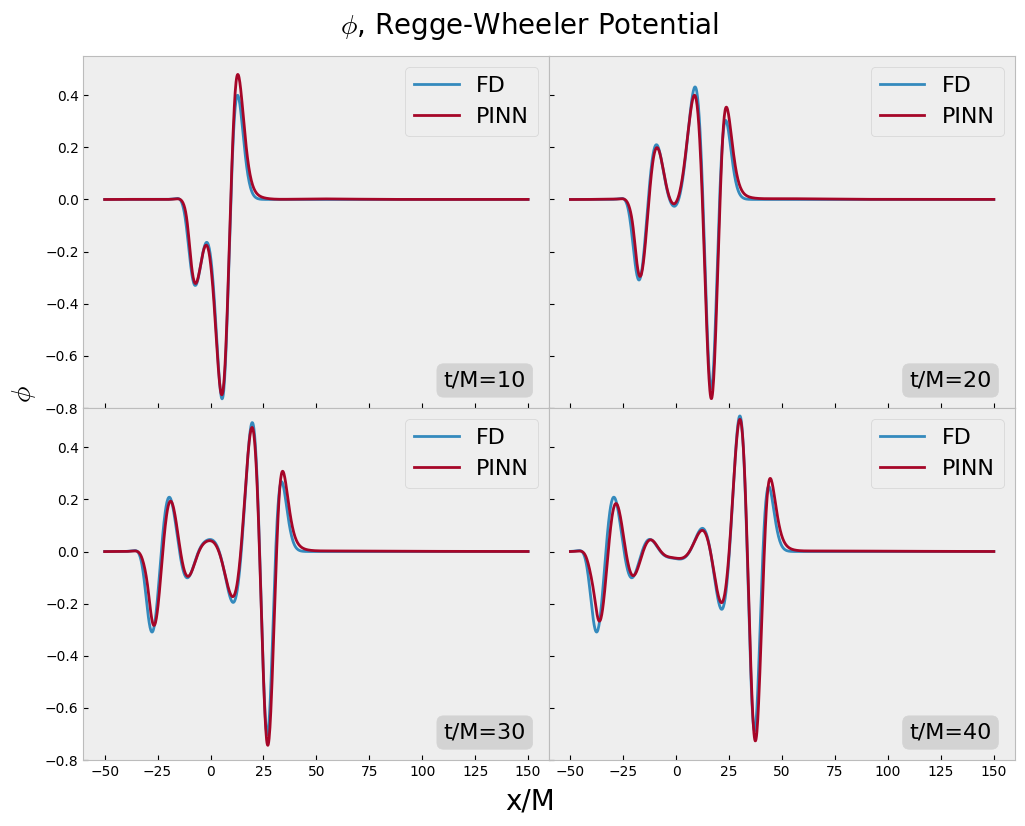}
    \caption{\label{RWPotPlot} Snapshots of the solutions for the Regge-Wheeler Potential.}
\end{figure}

\begin{figure}[hbt!]
    \includegraphics[width=1\textwidth]{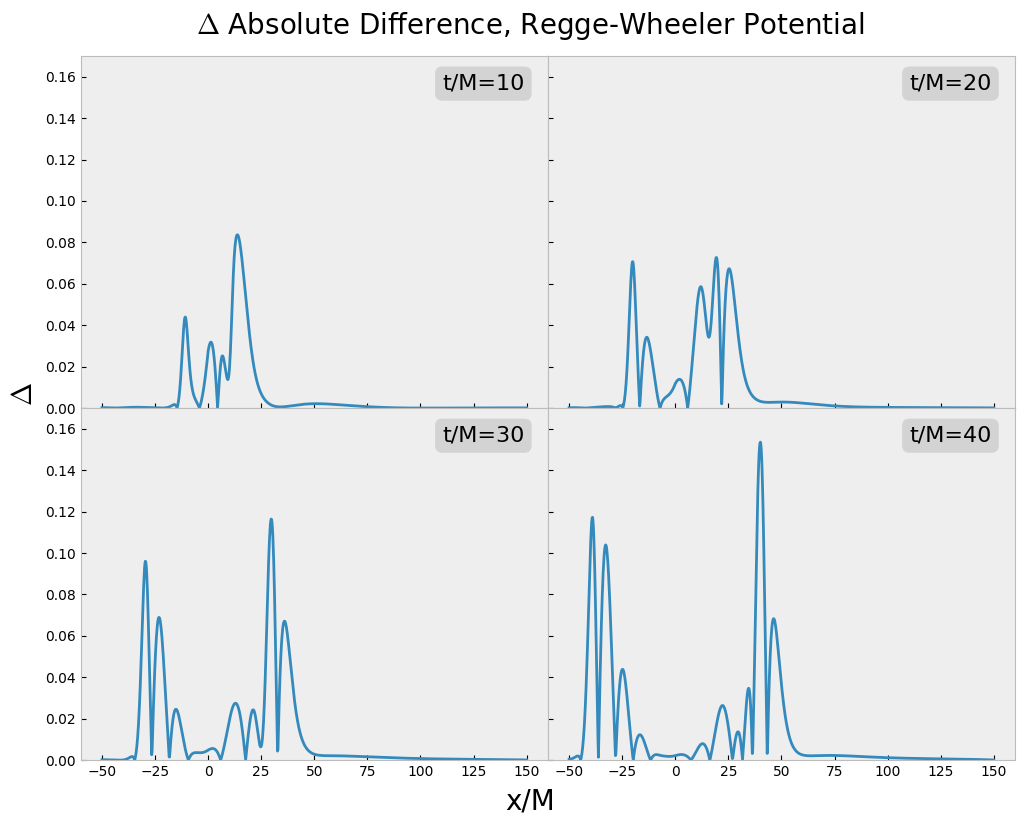}
    \caption{\label{RWPotDiffPlot}Absolute difference between the \FD{} and \pinn{} solutions (Regge-Wheeler Potential).}
\end{figure}

\begin{figure}[hbt!]
    \includegraphics[width=1\textwidth]{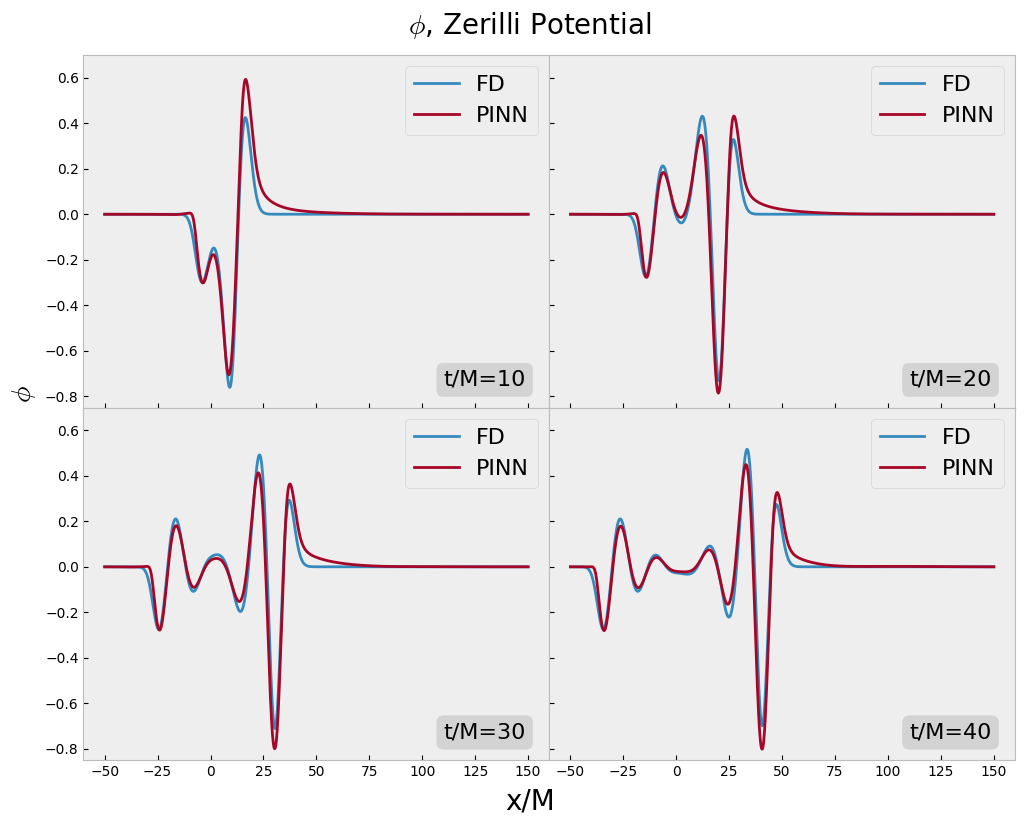}
    \caption{\label{ZPotPlot}Snapshots of the solutions for the Zerilli Potential.}
\end{figure}

\begin{figure}[hbt!]
    \includegraphics[width=1\textwidth]{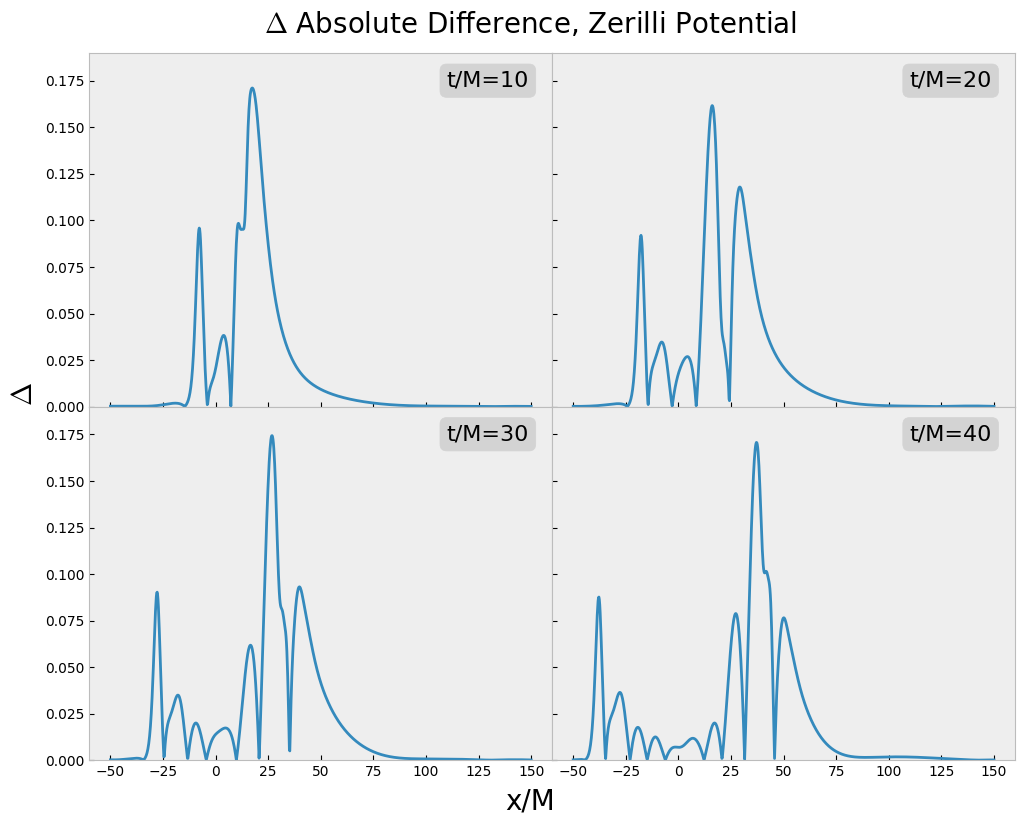}
    \caption{\label{ZPotDiffPlot}Absolute difference between the \FD{} and \pinn{} solutions (Zerilli Potential).}
\end{figure}

Figure~\ref{RWPotPlot} shows four snapshots of $\Phi(x,t)$ at times $t/M=10,20,30,\textnormal{ and }40$ for the case of a Regge-Wheeler potential. In red is the solution from \FD{} and in blue that from \pinn{}. In Figure~\ref{RWPotDiffPlot}, we show the absolute value difference between the \FD{} and \pinn{} solutions at the corresponding times. The most significant differences are observed at the leading edge of the pulse. For the case of the Zerilli potential, Fig.~\ref{ZPotPlot} shows snapshots in the evolution and Fig.~\ref{ZPotDiffPlot} the differences between \FD{} and \pinn{.} Here again, the most prominent differences are observed at the leading edge.


The main reason for the differences is because \pinn{s} for linear \PDE{s} suffer from a phase problem. Since in Eq.~(\ref{eq:PDE}) the potential is localized around $x=1$, to a good approximation, in most of the computational domain, we are solving the wave equation. If $\Phi(x,t)$ is a solution, so $\Phi(x,t+\alpha)$ with $\alpha$ a constant. Thus, \pinn{} may approach a solution with some associated phase instead of the true solution. To avoid this pitfall, we set the weight of initial velocity loss to 100, forcing \pinn{} to minimize the phase $\alpha$. 

\begin{figure}[hbt!]
    \includegraphics[width=1\textwidth]{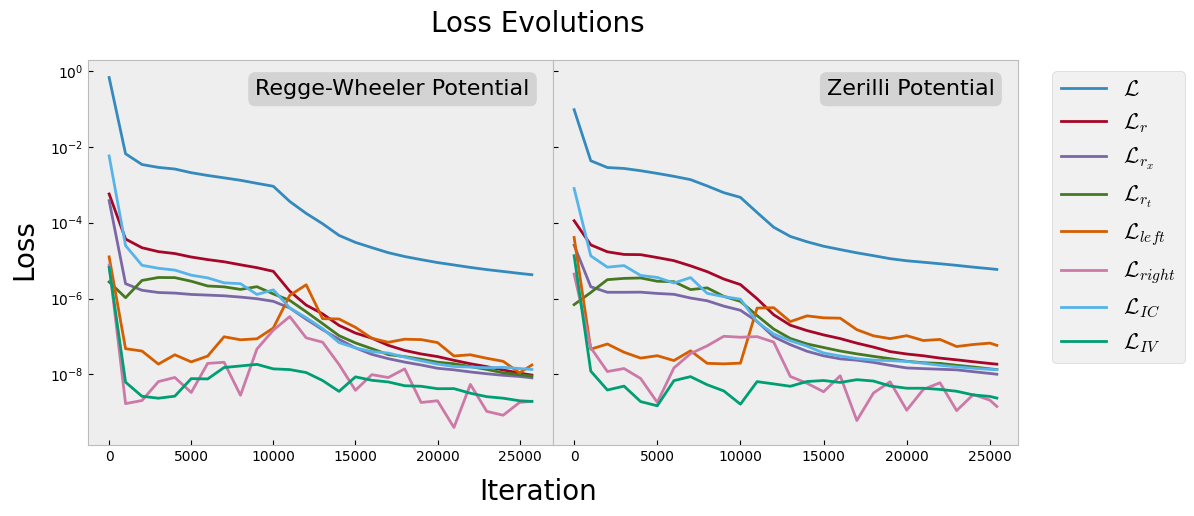}
    \caption{\label{LossPlot}Evolution of loss function and each of its components as a function of iterations.}
\end{figure}

In Figure~\ref{LossPlot}, we show the evolution of the total loss function and of each of its components as a function of iterations. 

\begin{figure*}[hbt!]
    \includegraphics[width=1\textwidth]{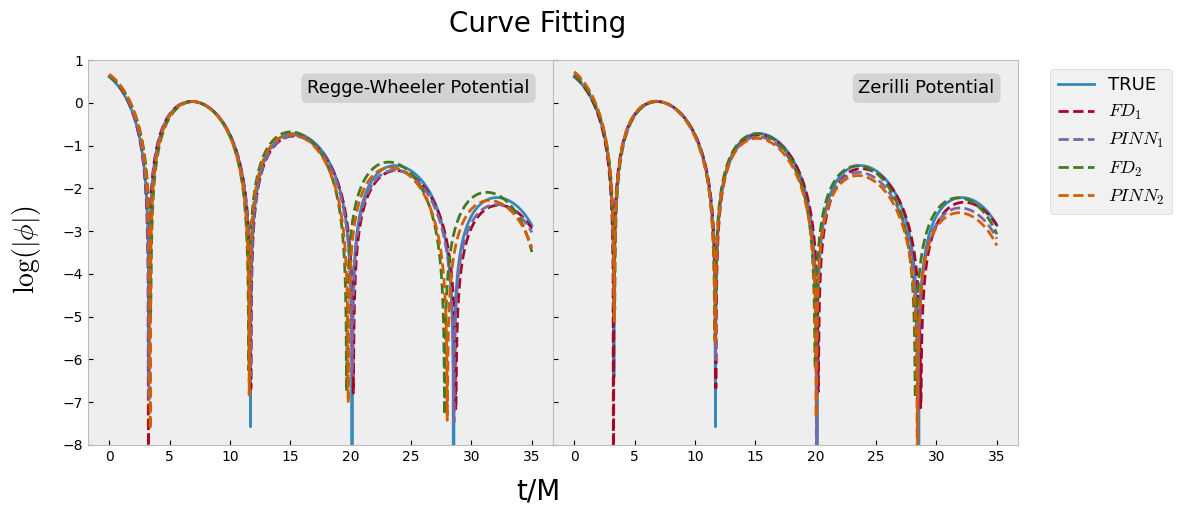}
    \caption{\label{CF}Curve Fittings for both Regge-Wheeler and Zerilli Potential.}
\end{figure*}

To calculate the \qnm{s}, we sampled the solution $\Phi(x,t)$ of Eq. \ref{eq:PDE} for both the Regge-Wheeler and Zerilli potentials at $x_q = 10\,M$. As mentioned before, at late times after the initial burst passes, the solution will behave as $\Phi(x_q,t) \propto e^{-t/\tau}\cos{(\omega t)}$. Figure~\ref{CF} shows $\Phi(x_q,t)$ for $\ell = 2$ from \pinn{s}, \FD{s}, and the solution using the values for $\omega$ and $\tau$ from Table~\ref{table:QNM}. The left panel is for the Regge-Wheeler potential, and the right panel is for Zerilli. 

We followed two approaches to calculate $\omega$ and $\tau$ from the \pinn{} and \FD{} solutions. In method 1, a Fourier transformation was applied to $\Phi$ to find the frequency $\omega$. For the decay time, $\tau$ was found by fitting a straight line passing through the local maxima of the log plots in Fig.~\ref{CF}. In method 2, we applied a direct curve fit $e^{-t/\tau}\cos{(\omega t)}$ to the data. The results from both fits are given in Table~\ref{ResultTableCF}.

\begin{table}[htbp]
    \centering
    \begin{tabular}{|c|c|c|}
        \hline
                        & $\boldsymbol{\tau /M}$ & $\boldsymbol{\omega\,M}$  \\ \hline
        & \multicolumn{2}{c|}{\textbf{Regge-Wheeler}} \\ \hline
        \textbf{FD$_1$}   & $10.488 \, (6.70)$ & $0.370 \, (0.90)$   \\ \hline
        \textbf{FD$_2$}   & $11.489 \, (2.21)$ & $0.387 \, (3.68)$ \\ \hline
        \textbf{PINN$_1$} & $10.474 \, (6.82)$ & $0.373 \, (0.11)$  \\ \hline
        \textbf{PINN$_2$} & $10.636 \, (5.39)$ & $0.383 \, (2.51)$ \\ \hline
                & \multicolumn{2}{c|}{\textbf{Zerilli}} \\ \hline
        \textbf{FD$_1$}   & $10.804 \, (3.89)$ & $0.370 \, (0.90)$   \\ \hline
        \textbf{FD$_2$}   & $11.073 \, (1.49)$ & $0.378 \, (1.26)$  \\ \hline
        \textbf{PINN$_1$} & $10.089 \, (10.25)$ & $0.375 \, (0.29)$  \\ \hline
        \textbf{PINN$_2$} & $9.620 \, (14.42)$ & $0.376 \, (0.52)$ \\ \hline
    \end{tabular}
    \caption{\qnm{} parameters $\omega$ and $\tau$ for $\ell = 2$. In parenthesis are the percentage error. The subindices denote the method used. For reference, the values from the literature are 
    $\tau = 11.241\,M $ and $\omega =  0.374 / M$ for the $\ell = 2$ mode.}
    \label{ResultTableCF}
\end{table}

\section{Conclusions and Future Directions}\label{Conclusions}

The surge in popularity of neural networks in data science has also triggered applications in which input from physics is used to accelerate discovery and make more accurate predictions. In the present study, we have applied \pinn{s} to solve the Zerilli and the Regge-Wheeler equations in the time domain to compute the \qnm{s} of a Schwarzschild \bh{}. At the core of the problem is the construction of a loss function with terms capturing the physics of the problem, e.g., residuals of the equations, boundary conditions, and initial data. To estimate the accuracy and efficiency of \pinn{s} for this problem, we compare the extracted \qnm{s} with those obtained with \FD{} methods. 
The results show that the \pinn{} approach is competitive but not more accurate than \FD{} methods. 
At the same time, our results show that \pinn{s} solve the equations under consideration with only a fraction of collocation points than \FD{}. This supports the view that as the dimensionality of the problem increases, \pinn{s} become more efficient as a tool to explore parameter space and guide where to carry out higher accuracy simulations with other methods.

Apart from using a gradient-enhanced \pinn{}, here are some improvements to mitigate errors that we will consider in future work:
\setlist{nolistsep}
\begin{itemize}[leftmargin=*,noitemsep]\label{Improvements}
  \item \emph{Residual-based adaptive sampling: }Wu et al. \cite{WU2023115671} experimented with different non-adaptive and residual-based adaptive for \pinn{} and developed structures of two adaptive samplings:  residual-based adaptive distribution (RAD) and residual-based adaptive refinement with distribution (RAR-D). RAD analyzes the current \pinn{} performance and creates a new batch of training points according to the distribution of residual. On the other hand, RAR-D analyzes the current \pinn{} performance and appends $m$ residual points with the highest residuals to the training domain. Compared to RAD, RAR-D utilizes less computing resources. 
  \item \emph{Learning rate annealing: }In some cases, \pinn{} fails because of having stiff loss gradients that hinder the progress of the gradient descent method because of it falling in a limit cycle or becoming unstable \cite{Wang2021}. Wang et al. \cite{Wang2021} proposed a learning rate annealing method that updates the weights of the loss terms at each iteration of the gradient descent, allowing the gradient to be more gentle and relaxing the interactions between competing loss terms. 
  \item \emph{Improved fully-connected neural architecture: } The accuracy of the \pinn{} can be improved by feeding it the prior knowledge about the physical system, for example, symmetry, conservation laws, etc. However, many physical systems are poorly understood, obstructing the suggested exploitation. However, extending the architecture of the \pinn{} to include general symmetry can improve its accuracy. Wang et al. \cite{Wang2021} suggested one such extension: introducing two transformer networks that account for the multiplicative interaction between input dimensions and augment the hidden states with residual connections. This extension provides about a ten-fold improvement compared to traditional \pinn{} \cite{Wang2021}. 
  \item \emph{Incorporating causality: } In some cases, particularly with chaotic cases, PINN tends to approach incorrect solutions because of it being inconsiderate of causality. Wang et al. \cite{Wang2022Causality} proposed an algorithm to incorporate the temporal causality in PINN. The algorithm gives weights to the residual loss terms at different times and updates them concerning them at times before that particular time. For weight updates, it uses exponential decay, so the residual loss term at any particular time will only be considered if the residual loss terms before that time are minimized enough. This idea can be extended to spatial causality, so any residual point is only affected by the points inside the lower half of the light cone.
  \item \emph{Multi-scale Fourier feature embeddings: }Wave equations exhibit high frequencies, and \pinn{s} usually fail to capture the high-frequency nature of the solutions. Wang et al. \cite{WANG2021113938} examined the high-frequency and multi-scale problems through the lens of Neural Tangent Kernel (NTK). After analyzing the spectral bias, they presented spatio-temporal multi-scale Fourier feature architecture. The architecture allows the network to learn frequencies determined by problem-dependent input parameter $\sigma$. 
\end{itemize}
\FloatBarrier

\section{Acknowledgments}

This work is supported by NSF grants PHY-2114582 and PHY-2207780. 

\bibliographystyle{elsarticle-num}
\bibliography{refs}

\end{document}